# BALLISTIC 3-PORT INTERFEROMETRIC LOGIC GATES IN THE QUANTUM HALL REGIME

A. Radu, S. Iftimie, D. Dragoman*

Faculty of Physics, University of Bucharest, P.O. Box MG-11,

077125 Bucharest-Magurele, Romania

**Abstract**

We show that a 3-port interferometer, defined by quantum point contacts in a two-dimensional electron gas, and working in the quantum Hall regime, can implement different logic gates at each port depending on the energy of charge carriers and the operating conditions. In all cases the input logic states are encoded in the potentials applied on the quantum point contacts, while the output logic states are encoded in the values of the overall transmission coefficients at each port. In addition, the output of such compact configurations can be reprogrammed by combining two 3-port interferometers.

___________________________________________________________________________

* Corresponding author: danieladragoman@yahoo.com

**Introduction**

The quest for miniaturization of integrated circuits, brought forward by efforts to go beyond Moore's law [1], lead to development of an increasing number of electronic and optoelectronic devices that rely on low-dimensional nanostructures, in particular on two-dimensional electron gases (2DEGs). As a consequence, quantum mechanics has become not only an indispensable tool to describe charge carrier transport in such structures but, as quantum computing [2], has evolved into an alternative to classical computing techniques. Presently, many proposals of physical systems able to implement quantum logic gates and circuits have been put forward, and remarkable progress has been registered in their implementation [3], although there are still considerable technological difficulties to be overcome. In particular, solid-state quantum computing systems have been thoroughly investigated, charge- [4,5], spin- [6,7] or superconductor-based [8] configurations being mainly proposed as quantum logic gates. A different route for implementing logic operations in solid-state systems involves configurations in materials with long mean-free-paths of charge carriers, such as graphene [9-11] or 2DEGs [12] based on interference between quantum wavefunctions traversing different paths. This last category includes also logic gates based on quantum interference between edge states in electron interferometers defined by quantum point contacts (QPCs) [13,14]. Such few- or single-channel electron interferometers working in the quantum Hall effect regime have the advantage of high sensitivity and visibility [15], as well as of a well-establish, even analytical [13,16,17], modeling of propagating single-particle wavefunctions or wavepackets. Moreover, conditional operations involving logic gates implemented by electronic interferometers based on edge states could be envisaged taking into account the Coulomb interaction of different interferometers [14,18,19] or the coupling of different cavities by edge states [20].

In this paper we propose a logic gate configuration that has the advantage of compactness, which is essential for preserving the overall coherence of the electron wavefunction along a succession of logic gates, and ease of conditional reprogramming, which is important for implementing reconfigurable logic circuits. The proposed configuration is a 3-port interferometer, patterned on a 2DEG in the integer quantum Hall regime, the charge carriers propagating along edge state channels. The input logic states are encoded on the potentials applied on the QPCs that define the interferometer, while the output logic states are determined by the overall wavefunction transmission coefficient at each port. In this configuration each output is a linear combination of the 3 possible inputs/edge state sources, which determine the operating condition of the gate. As such, these logic gates have the distinctive feature of being able to be combined/reprogrammed in a simple manner, the output of one gate defining the operating condition of the next logic gate. Moreover, because different logic gates can be implemented at each port, the device offers the unique possibility to process in parallel different algorithms.

**3-port logic gates based on edge state channels**

We model the operation of the proposed device considering ballistic charge carrier transport in a quantum Hall cavity/interferometer defined by three QPCs and patterned in a 2DEG. Throughout this paper we assume spin-independent scattering of charge carriers by the potential barriers. More precisely, the 2DEG is supposed to be confined along the $z$ direction, i.e. it extends in the $(x,y)$ plane, and a uniform magnetic field $B$ is applied along $z$, such that the vector potential can be chosen as $A = (0, Bx, 0)$. Then, as discussed in [13], the single-particle Hamiltonian describing the propagation of an electron with effective mass $m$ and wavevector components along the $x$ and $y$ directions $k_x$ and $k_y$, respectively, can be written as

$$H = \frac{\omega_d^2}{\omega_t^2}\frac{\hbar^2 k_y^2}{2m} + \frac{\hbar^2 k_x^2}{2m} + \frac{m\omega_t^2}{2}(x-x_0)^2 \tag{1}$$

where $\omega_d$ is the frequency associated to the transverse confining potential along the $x$ direction, $\omega_t = \sqrt{\omega_d^2 + \omega_c^2}$ with $\omega_c = eB/m$ the cyclotron frequency, and $x_0 = \hbar k_y \omega_c /(\omega_t^2 m)$. The dispersion relation of the quantized energy levels is then given by

$$E_{n,k_y} = \hbar^2 k_y^2 \frac{\omega_d^2}{2m\omega_t^2} + \hbar\omega_t\left(n+\frac{1}{2}\right), \tag{2}$$

two edge states, i.e. two-split channels localized on the opposite edges $\pm x_0$, corresponding to wavevector components $\pm k_y$, being allowed to propagate at a fixed energy value. We consider in the following that only the lowest subband, with $n = 0$, is occupied, limiting the range of energies $E$ between $\hbar\omega_t/2$ and $3\hbar\omega_t/2$. It should be noted that in a 2DEG the energy (Fermi energy) of charge carriers can be tuned by applying gate voltages (see, for instance, [21]).

The schematic representation of the 3-port logic gate is represented in Fig. 1. We model the action of each of the three QPCs, situated at $x = x_i$, $i = 1,2,3$, by a delta-function potential $U_i(x_i) = (\hbar^2/2m)\gamma_i\delta(x_i)$. If the $i$th QPC separates regions in which the wavefunction has the same wavevector, denoted by $k$, the incident wavefunction is partly reflected, with reflection coefficient $r_i = \gamma_i/(2ik-\gamma_i)$, and partly transmitted with transmission coefficient $t_i = 2ik/(2ik-\gamma_i)$, the strength of the QPC depending on the applied potential $U_i$.

The output vector of the 3-port interferometric device, $out^T = (out1, out2, out3)$, where the superscript $T$ denotes transposition, depends linearly on the input vector $in^T = (in1, in2, in3)$ as $out^T = M \cdot in^T$ or, written on components,

$$\begin{pmatrix} out1 \\ out2 \\ out3 \end{pmatrix} = \begin{pmatrix} M_{11} & M_{12} & M_{13} \\ M_{21} & M_{22} & M_{23} \\ M_{31} & M_{32} & M_{33} \end{pmatrix} \begin{pmatrix} in1 \\ in2 \\ in3 \end{pmatrix} \quad (3)$$

To find out the elements of the matrix $M$, let us suppose first that $in^T = (in1, 0, 0)$, i.e. that there is only one input, through QPC1. Then,

$$out1 = in1\left[r_1 + t_1\left(t_1 r_2 r_3 e^{i\Phi}\right) + t_1\left(t_1 r_2 r_3 e^{i\Phi}\right)\left(r_1 r_2 r_3 e^{i\Phi}\right) + ....\right] = in1\left[r_1 + \frac{t_1^2 r_2 r_3 e^{i\Phi}}{1 - r_1 r_2 r_3 e^{i\Phi}}\right] \quad (4a)$$

$$out2 = in1\left[\left(t_1 e^{i\phi_{12}}\right)t_2 + \left(t_1 e^{i\phi_{12}}\right)\left(r_2 r_3 r_1 e^{i\Phi}\right)t_2 + ....\right] = in1\frac{t_1 t_2 e^{i\phi_{12}}}{1 - r_1 r_2 r_3 e^{i\Phi}} \quad (4b)$$

$$out3 = in1\left[\left(t_1 e^{i\phi_{12}} r_2 e^{i\phi_{23}}\right)t_3 + \left(t_1 e^{i\phi_{12}} r_2 e^{i\phi_{23}}\right)\left(r_1 r_2 r_3 e^{i\Phi}\right)t_3 + ....\right] = in1\frac{t_1 r_2 t_3 e^{i(\phi_{12}+\phi_{23})}}{1 - r_1 r_2 r_3 e^{i\Phi}} \quad (4c)$$

where $\phi_{ij} = kL_{ij}$ denotes the phase acquired by the wavefunction when propagating from port $i$ to port $j$, separated by the distance $L_{ij}$, and $\Phi = \phi_{12} + \phi_{23} + \phi_{31}$. Similar calculations of outputs for inputs at ports 2 and 3, defined by QPC2 and QPC3, lead us to find the elements of matrix $M$ as:

$$M_{11} = r_1 + \frac{t_1^2 r_2 r_3 \exp(i\Phi)}{\Delta}, \quad M_{21} = \frac{t_1 t_2 \exp(i\phi_{12})}{\Delta}, \quad M_{31} = \frac{t_1 r_2 t_3 \exp[i(\phi_{12}+\phi_{23})]}{\Delta} \quad (5a)$$

$$M_{12} = \frac{t_2 r_3 t_1 \exp[i(\phi_{23} + \phi_{31})]}{\Delta}, \quad M_{22} = r_2 + \frac{t_2^2 r_3 r_1 \exp(i\Phi)}{\Delta}, \quad M_{32} = \frac{t_2 t_3 \exp(i\phi_{23})}{\Delta} \quad (5b)$$

$$M_{13} = \frac{t_3 t_1 \exp(i\phi_{31})}{\Delta}, \quad M_{23} = \frac{t_3 r_1 t_2 \exp[i(\phi_{31} + \phi_{12})]}{\Delta}, \quad M_{33} = r_3 + \frac{t_3^2 r_1 r_2 \exp(i\Phi)}{\Delta} \quad (5c)$$

where $\Delta = 1 - r_1 r_2 r_3 \exp(i\Phi)$. The transmission coefficient at port $i$ is then calculated by estimating the square modulus of *out_i*, $i = 1, 2, 3$.

Formulas (5) allow us to estimate the output wavefunctions at each of the 3 ports depending on the input wavefunctions. To correlate the outputs with the operation of logic gates, we must first specify the manner of encoding the logic values of both inputs and outputs. As in [13], we encode the input logic values in the potential applied on each of the three QPCs, which can be either $V_0 = 0$, corresponding to the logic value 0, or $V_1 = 0.5$ eV, associated to the logic value 1. The input logic states are then $(U_1, U_2, U_3)$, with $U_i$, $i = 1, 2, 3$, equal to $V_0$ or $V_1$. The input vector *in* is to be regarded as parameter, while the output logic states are encoded in the values of the transmission coefficients *T* at each port, as exemplified below.

Figure 2 illustrates the energy dependences of the three outputs when $in^T = (1,0,0)$, the input logic states encoded as $(U_1, U_2, U_3)$ being specified in the legend. Because several curves overlap, a three-dimensional representation is shown as inset. In all figures in this paper the relevant energy range is chosen as discussed above, the parameters used for simulations being $m = 0.067 m_0$ with $m_0$ the free electron mass (as for electrons in the GaAs/AlAs 2DEG system), $\hbar\omega_d = \hbar\omega_c = 4$ meV, and $L_{12} = L_{23} = L_{31} = 100$ nm. From Fig. 2 it follows that the transmission coefficients *T* at all ports have a smooth variation with energy except for the input logic state (111), for which rapid variations of the transmission occur for specific resonant energies denoted as $E_r$ (a similar behavior was reported in [13]).

Because each output wavefunction could be separately detected, the configuration presented in Fig. 1 offers the opportunity to implement different logic gates on each of the three outputs, and even different logic gates on a specific output depending on the energy of charge carriers and the threshold imposed to the transmission coefficient to distinguish between logic states 0 and 1. We will refer in the following to this threshold value as $T_{th}$, transmission coefficients $T > T_{th}$ being associated to the output logic value 1, while the output logic value 0 is encoded in $T < T_{th}$. More precisely, looking at the three outputs in Fig. 2, for $in^T = (1,0,0)$, we observe that:

I. For *out*1, considering $T_{th} = 0.5$, we can implement, for example:

    a) the NOT gate for $U_1$ at resonance (at $E_r$), if $(U_2, U_3) = (11)$

    b) the AND gate for $U_2$ and $U_3$ at all energies, if $U_1 = 0$

    c) the OR gate for $U_1$ and $U_3$ outside resonance, if $U_2 = 1$

    d) the OR gate for $U_1$ and $U_2$ outside resonance, if $U_3 = 1$

    e) the identity operation for $U_1$ at all electron energies, if $(U_2, U_3) = (10)$, (00), or (01)

    f) the Set 1 gate (sets the output logic state to 1), for $U_1 = 1$ outside resonance, irrespective of the logic values of $U_2$, $U_3$

    c) the CCNOT (Toffoli gate), which changes the value of the target bit when both control bits are 1, at $E_r$, if $U_1$ is the target bit and $U_2$ and $U_3$ are control bits

II. For *out*2, considering $T_{th} = 0.3$, we can implement, for example:

    a) the NOT gate for $U_1$ at all energies, if $(U_2, U_3) = (01)$ or (00)

    b) the modified CNOT gate (it changes the value of the target bit if the control bit is 0) at $E_r$, if $U_3 = 1$, with $U_1$ the control bit and $U_2$ the target bit

    c) the AND gate for $U_2$ and $U_3$ at $E_r$, when $U_1 = 1$

    d) the AND gate for $U_1$ and $U_3$ at $E_r$, when $U_2 = 1$

    e) the NOR gate for $U_1$ and $U_2$ at all energies, when $U_3 = 0$

e) the identity operation for $U_1$ at $E_r$, if $(U_2,U_3) = (11)$

III. For *out*3, considering $T_{th} = 0.3$, we can implement, for example:

   a) the AND gate for $U_1$ and $U_2$ at $E_r$, when $U_3 = 1$

   b) the AND gate for $U_2$ and $U_3$ at $E_r$, when $U_1 = 1$

   c) the NOT gate for $U_1$ at all energies, when $(U_2,U_3) = (10)$

   d) the Erase gate (sets the output logic state to 0) at all energies, if $U_2 = 0$, irrespective of the logic values of $U_1$ and $U_3$

   e) the NOR gate for $U_1$ and $U_3$ outside resonance, if $U_2 = 1$

   f) the identity operator for $U_1$ at $E_r$, if $(U_2,U_3) = (11)$

These examples show that one can implement different logic gates at different outputs of the 3-port interferometric configuration in Fig. 1. As such, this configuration allows a parallel processing of different quantum gates and algorithms starting from the same input states.

The energy dependences of the three outputs when $in^T = (1,1,0)$ for different input logic states given in the legend are shown in Fig. 3; the three-dimensional representation in the inset helps identify the overlapping curves. Note that in this situation the maximum transmission value at each port can be higher than 1 (but smaller than 2), since there are two inputs, each with magnitude 1. Again, the output transmission values at each port vary in a smooth manner with energy, except the case when $(U_1,U_2,U_3) = (111)$. However, the energy dependence of $T$ is not monotonous, which implies that the logic gates that could be implemented at each port depend also on the energy values of charge carriers. For instance,

I. For *out*1,

i) at resonant energies $E_r$, considering $T_{th} = 0.5$, we can implement, for example:

   a) the OR gate for $U_1$ and $U_2$, if $U_3 = 0$

   b) the NAND gate for $U_1$ and $U_3$, if $U_2 = 1$

c) the NAND gate for $U_2$ and $U_3$, if $U_1 = 1$

   d) the Set 1 gate, when $(U_2, U_3) = (01)$ or $(10)$

ii) outside resonance, considering $T_{th} = 0.3$, we can implement, for example:

   a) the OR gate for $U_1$ and $U_3$, irrespective of $U_2$

   b) the Set 1 gate, when $U_1 = 1$ or $U_3 = 1$

iii) at $E \cong 4.3$ meV, considering $T_{th} = 1.2$, we can implement, for example:

   a) the Erase gate if $U_1 = 1$, irrespective of $U_2$ and $U_3$

   b) the AND gate for $U_2$ and $U_3$, if $U_1 = 0$

iv) at $E \cong 5.7$ meV, considering $T_{th} = 1.25$, we can implement, for example:

   a) the Erase gate when $(U_1, U_2) = (00), (01)$ or $(11)$, irrespective of $U_3$

   b) the AND gate for $U_1$ and $U_3$, if $U_2 = 0$

v) at $E \cong 6$ meV, considering $T_{th} = 0.7$, we can implement, for example:

   a) the NOR gate for $U_1$ and $U_3$, if $U_2 = 0$

   b) the Set 1 gate when $U_1 = 1$, irrespective of $U_2$ and $U_3$

   c) the identity operation for $U_1$, if $(U_2, U_3) = (00), (10),$ or $(11)$

vi) at $E \cong 7.7$ meV, considering $T_{th} = 1.2$, we can implement, for example:

   a) the Erase gate if $U_3 = 0$, irrespective of $U_1$ and $U_2$

   b) the AND gate for $U_1$ and $U_3$, if $U_2 = 0$

   c) the AND gate for $U_2$ and $U_3$, if $U_1 = 0$

   d) the CNOT gate when $U_3 = 1$, with $U_1$ the target bit and $U_2$ the control bit

II. For *out2*,

i) at resonant energies $E_r$, considering $T_{th} = 0.5$, we can implement, for example:

   a) the NOR gate for $U_1$ and $U_2$, if $U_3 = 1$

   b) the NOR gate for $U_1$ and $U_3$, if $U_2 = 1$

   c) the CNOT if $U_1 = 1$, with $U_2$ target bit and $U_3$ control bit

    d) the Set 1 gate, when $U_1 = 0$, irrespective of $U_2$ and $U_3$

    e) the NOT gate for $U_1$, if $(U_2,U_3) = (00)$ or $(11)$

ii) outside resonance, considering $T_{th} = 0.3$, we can implement, for example:

    a) the OR gate for $U_2$ and $U_3$, if $U_1 = 0$

    b) the Set 1 gate, when $U_1 = 0$, $U_2 = 1$ or $U_3 = 1$

iii) at $E \cong 4.8$ meV, considering $T_{th} = 1.2$, we can implement, for example:

    a) the Erase gate if $U_1 = 0$, $U_2 = 1$, or $U_3 = 0$

    b) the AND gate for $U_1$ and $U_3$, if $U_2 = 0$

iv) at $E \cong 5.7$ meV, considering $T_{th} = 0.8$, we can implement, for example:

    a) the Set 1 gate if $U_1 = 0$ or $U_2 = 1$

v) at $E \cong 6$ meV, considering $T_{th} = 1.2$, we can implement, for example:

    a) the Erase gate if $U_1 = 1$ or $U_2 = 0$

    b) the NOT gate for $U_1$, if $(U_2,U_3) = (10)$ or $(11)$

III. For *out3*,

i) outside resonance, considering $T_{th} = 0.5$, we can implement, for example:

    a) the NOR gate for $U_2$ and $U_3$, if $U_1 = 1$

    b) the NOT gate for $U_3$, if $(U_1,U_2) = (00)$, $(01)$ or $(10)$

ii) outside resonance, $E \cong 5\text{-}7$ meV, considering $T_{th} = 0.8$, we can implement, for example:

    a) the NOR gate for $U_2$ and $U_3$, irrespective of $U_1$

iii) outside resonance, $E \cong 4\text{-}5.2$ meV, with $T_{th} = 0.8$, we can implement, for example:

    a) the NOR gate for $U_1$ and $U_3$, if $U_2 = 1$

    b) the NOR gate for $U_2$ and $U_3$, if $U_1 = 1$

    c) the NAND gate for $U_1$ and $U_2$, if $U_3 = 0$

    d) the NOT gate for $U_3$, if $(U_1,U_2) = (00)$ or $(10)$

    e) the Erase gate if $U_3 = 1$, irrespective of $U_1$ and $U_2$

iv) at resonance, $E \cong 5.2$ meV, with $T_{th} = 0.8$, we can implement, for example:

a) modified CNOT gate if $U_1 = 1$, with $U_3$ the target bit and $U_2$ the control bit

These examples show that there is a large number of logic gates that can be implemented at each port for certain energies and transmission threshold values.

For the sake of completeness, we show in Fig. 4 the output transmission values at the three ports when $in^T = (1,1,1)$, for different input logic states specified in the legend. The overlapping curves can be identified using the three-dimensional representation in the inset. As opposed to the situations investigated above, no resonant energies exist and the three outputs have similar energy-dependent transmission but for different input logic states. This behavior is to be expected from the symmetry of the system. In a similar to the cases discussed above, a correlation with possible logic operations for particular charge carrier energies and $T_{th}$ values can be carried out in this case, as well. However, a further discussion of this issue would unnecessarily lengthen the present manuscript.

More important for practical applications is to illustrate how to combine two such 3-port interferometric devices. A schematic representation of such a configuration is shown in Fig. 5. The first device/ring (with ports 1, 2 and 3) operates conditionally to the output of the second device/ring (with ports 1', 2' and 3'). More precisely, for a combined system with two inputs, say, at ports 1 and 1', if the output at port 2' of the second device/ring is high when operating for $in^T = (1,0,0)$, in particular if the transmission $T = 1$, the first device operates in the $in^T = (1,1,0)$ regime. Otherwise, if the transmission at port 2' is zero, the output of the first device at a certain energy corresponds to the $in^T = (1,0,0)$ operating regime. In Fig. 5, QPC2 is identical to QPC2', which imposes a restriction/actually selects the logic gate implemented/detected at ports 1 or 3. To be more specific, if $(U_1',U_2',U_3') = (110)$, the transmission at port 2' is low (zero), and the first device with $U_2 = 1$ implements at port 3, for instance, the NOR operation for $U_1$ and $U_3$, outside resonance. If $(U_1',U_2',U_3') = (000)$ or

(001), the transmission at port 2' is high ($T = 1$), and the first device with $U_2 = 0$ implements at the same port the NOT operation for $U_3$, irrespective of $U_1$.

Note that the configuration in Fig. 5 is not an example of cascaded 3-port interferometers because of the way in which we have encoded the input logic states. To transform it into a cascaded configuration, the inputs *in* should not be regarded as parameters that specify the operating regime but as logic states instead of the applied potentials on QPCs. Then, $U_i$, $i = 1, 2, 3$, would become the parameters. Whether the simulation of this last configuration would not be difficult, its practical implementation would be much more complicated since to take advantage of all combinations of possible input logic states one would have to add another three gates for each 3-port interferometer, which would either allow or forbid the propagation/input of edge states at the ports.

**Conclusions**

We have shown that a 3-port interferometric configuration employing edge state channels can implement different logic gates at each port depending on the energy of charge carriers and the operating conditions. Another remarkable property of this compact logic device proposed in this paper is the possibility of reprogramming it by combining two 3-port interferometers. In the configurations studied in this paper the input logic states are encoded on the potentials applied on the QPCs that define the interferometer, while the output logic states are encoded in the values of the overall transmission coefficients at each port.

Note that throughout the paper we have considered the case of spin-independent scattering at potential barriers, which implies from a practical point of view either small-enough magnetic fields, such that the Zeeman splitting is not important, or that the spin-orbit coupling can be neglected, or that spin-polarized charge carriers are injected. The situation of spin-resolved propagation would complicate the interpretation of the results, not because charge carriers with opposite spins propagate along channels on the opposite sides of the 2DEG but, especially, because spin rotation takes place at quantum point contacts [14,22].

Throughout this paper we have limited the investigation to interferometers with three QPCs/ports, although configurations with a higher number of QPCs could be envisaged. The reason is that, besides being more difficult to fabricate, the increased complexity of interference patterns/transmission of outgoing wavefunctions render these configurations more difficult to operate, as well.

In conclusion, the simulation results obtained in this paper suggest that 3-port interferometers are complicated enough to offer a novel manner of implementing logic gates in parallel at each port and/or to reprogram these gates at will, without increasing significantly the technological challenges to fabricate them. As such, we are confident that these configurations will prove their utility in future computing designs.

**Acknowledgement**

This work was supported by a grant of Ministery of Research and Innovation, CNCS-UEFISCDI, project number PN-III-P4-ID-PCE-2016-0122, within PNCDI III

**References**

[1] M. M. Waldrop, More than Moore, Nature 530, 144-147 (2016).

[2] M. A. Nielsen, I. L. Chuang, *Quantum Computation and Quantum Information*, Cambridge Univ. Press, Cambridge, U. K. (2000).

[3] D. Castelvechi, Quantum computers ready to leap out of the lab, Nature 541, 9-10 (2017).

[4] H.-O. Li, G. Cao, G.-D. Yu, M. Xiao, G.-C. Guo, H.-W. Jiang, G.-P. Guo, Conditional rotation of two strongly coupled semiconductor charge qubits, Nature Commun. 6, 7681 (2015).

[5] L. Wang, T. Tu, B. Gong, C. Zhou, G.-C. Guo, Experimental realization of non-adiabatic universal quantum gates using geometric Landau-Zener-Stückelberg interferometry, Sci. Rep. 6, 19048 (2016).

[6] M. Veldhorst, H. G. J. Eenink. C. H. Yang, A. S. Dzurak, Silicon CMOS architecture for a spin-based quantum computers, Nature Commun. 8, 1766 (2017).

[7] L. M. K. Vandersypen, H. Bluhm, J. S. Clarke, A. S. Dzurak, R. Ishihara, A. Morello, D. J. Reilly, L. R. Schreiber, M. Veldhorst, Interfacing spin qubits in quantum dots and donors – hot, dense, and coherent, npj Quantum Information 3, 34 (2017).

[8] G. Wendin, Quantum information processing with superconducting circuits: a review, Rep. Prog. Phys. 80, 106001 (2017).

[9] D. Dragoman, M. Dragoman – Enhanced architectures for room-temperature reversible logic gates in graphene, Appl. Phys. Lett. 105, 113109 (2014).

[10] D. Dragoman, M. Dragoman – Quantum logic gates based on ballistic transport in graphene, J. Appl. Phys. 119, 094902 (2016).

[11] M. Dragoman, A. Dinescu, D. Dragoman – Wafer-scale fabrication and room-temperature experiments on graphene-based gates for quantum computation, IEEE Trans. Nanotechnology 17, 362-367 (2018).

[12] M. Yamamoto, S. Takada, C. Bäuerle, K. Watanabe, A. D. Wieck, S. Tarucha, Electrical control of a solid-state flying qubit, Nature Nanotechnol. 7, 247-251 (2012).

[13] S. Bellucci, P. Onorato, Quantum Hall Fabry-Pérot interferometer: logic gate responses, J. Appl. Phys. 108, 033710 (2010).

[14] W. Chen, Z.-Y. Xue, Z. D. Wang, R. Shen, D. Y. Xing, Quantum computing through electron propagation in the edge states of quantum spin Hall systems, Eur. Phys. J. B 87, 57 (2014).

[15] Y. Ji, Y. Chung, D. Sprinzak, M. Heiblum, D. Mahalu, H. Shtrikman, An electronic Mach-Zehnder interferometer, Nature 422, 415-418 (2003).

[16] A. Beggi, P. Bordone, F. Buscemi, A. Bertoni, Time-dependent simulation and analytical modelling of electronic Mach-Zehnder interferometry with edge-states wave packets, J. Phys.: Condens. Matter 27, 475301 (2015).

[17] S. Ngo Dinh, D. A. Bagrets, A. D. Mirlin, Analytically solvable model of an electronic Mach-Zehnder interferometer, Phys. Rev. B 87, 195433 (2013).

[18] E. Weisz, H. K. Choi, I. Sivan, M. Heiblum, Y. Gefen, D. Mahalu, V. Umansky, An electronic quantum eraser, Science 344, 1363-1366 (2014).

[19] J. Dressel, Y. Choi, A. N. Jordan, Measuring which-path information with coupled electronic Mach-Zehnder interferometers, Phys. Rev. B 85, 045320 (2012).

[20] J. Splettstoesser, S. Ol'khovskaya, M. Moskalets, M. Büttiker, Electron counting with a two-particle emitter, Phys. Rev. B 78, 205110 (2008).

[21] M. M. Uddin, H. W. Liu, K. F. Yang, K. Nagase, K. Sekine, C. K. Gaspe, T. D. Mishima, M. B. Santos, Y. Hirayama, Gate depletion of an InSb two-dimensional electron gas, Appl. Phys. Lett. 103, 123502 (2013).

[22] S. Bellucci, P. Onorato, Integer spin Hall effect in ballistic quantum wires, Phys. Rev. B 73, 045329 (2006).

**Figure captions**

Fig. 1  Schematic representation of the 3-port interferometric logic device

Fig. 2  Energy dependence of the transmission at (a) port 1, (b) port 2, and (c) port 3, in the device in Fig. 1 when $in^T = (1,0,0)$ for different input logic states specified in the legend. Insets: three-dimensional representation of the same dependence, to highlight overlapping curves

Fig. 3  Energy dependence of the transmission at (a) port 1, (b) port 2, and (c) port 3, in the device in Fig. 1 when $in^T = (1,1,0)$ for different input logic states specified in the legend. Insets: three-dimensional representation of the same dependence, to highlight overlapping curves

Fig. 4  Energy dependence of the transmission at (a) port 1, (b) port 2, and (c) port 3, in the device in Fig. 1 when $in^T = (1,1,1)$ for different input logic states specified in the legend. Insets: three-dimensional representation of the same dependence, to highlight overlapping curves

Fig. 5  Schematic representation of combined 3-port interferometric logic devices/rings

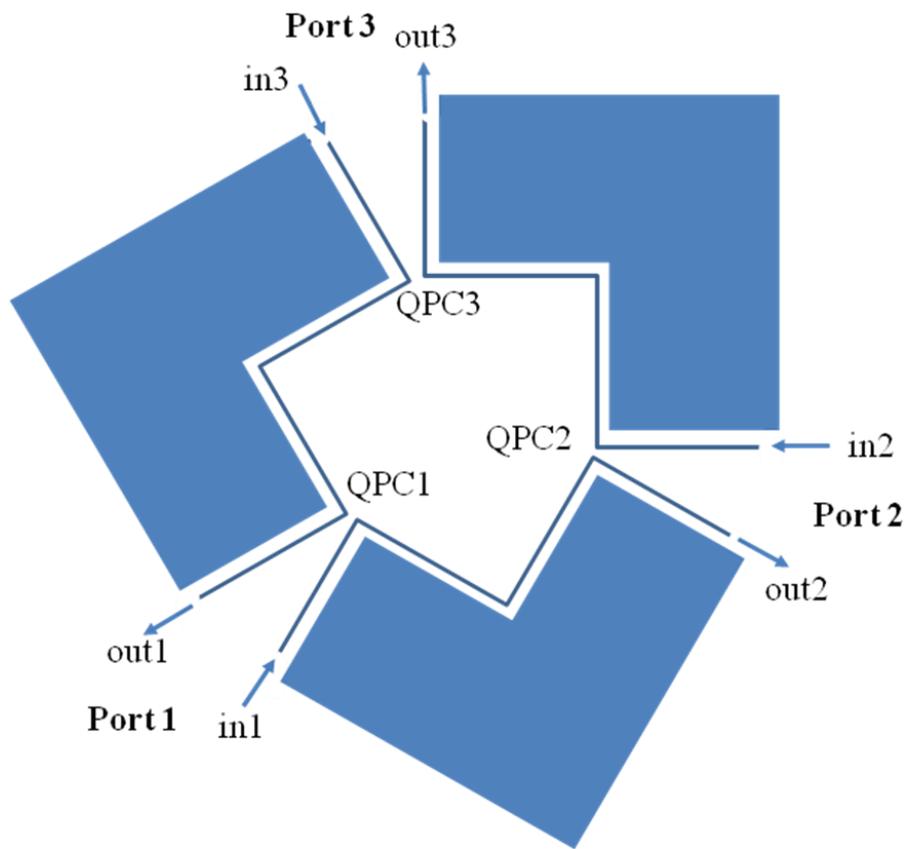

Fig. 1

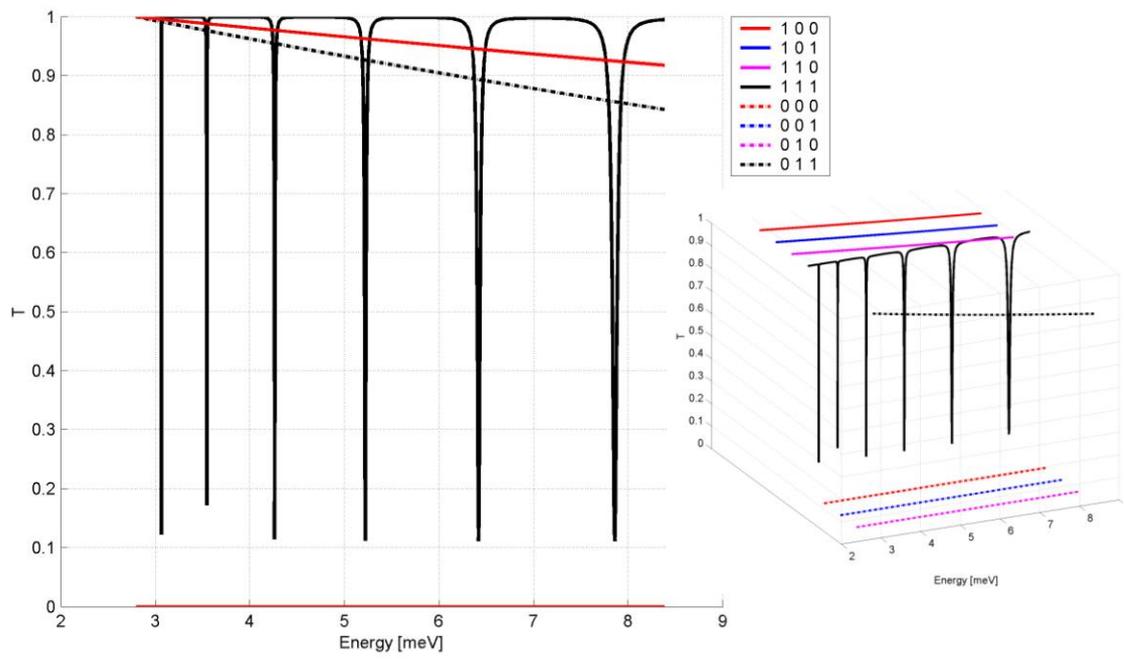

(a)

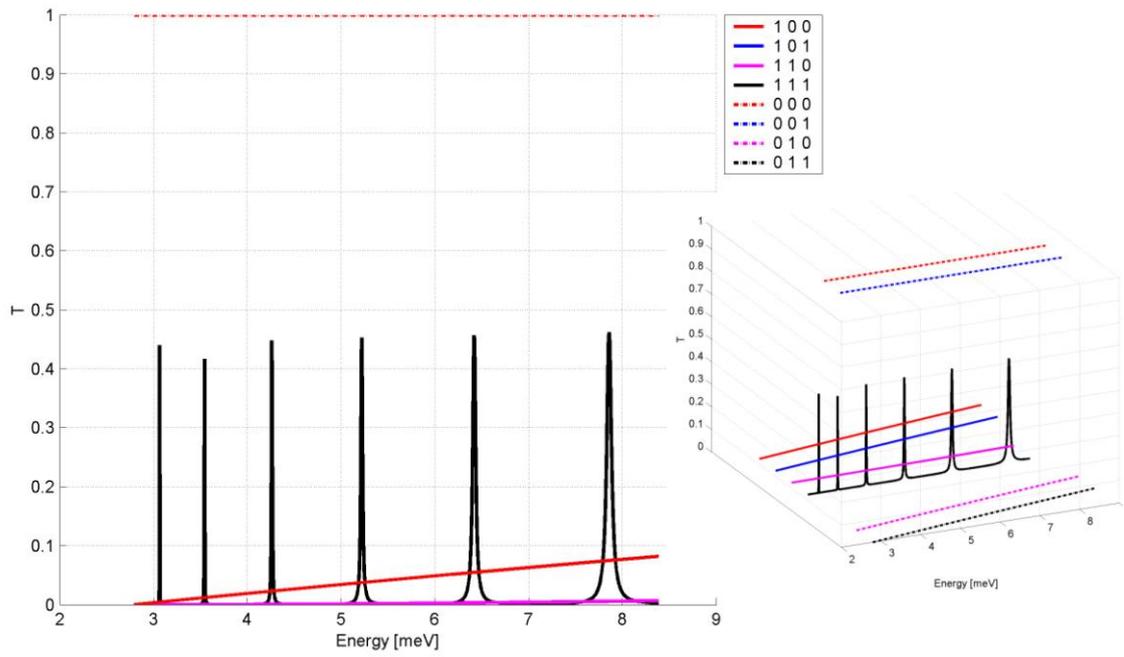

(b)

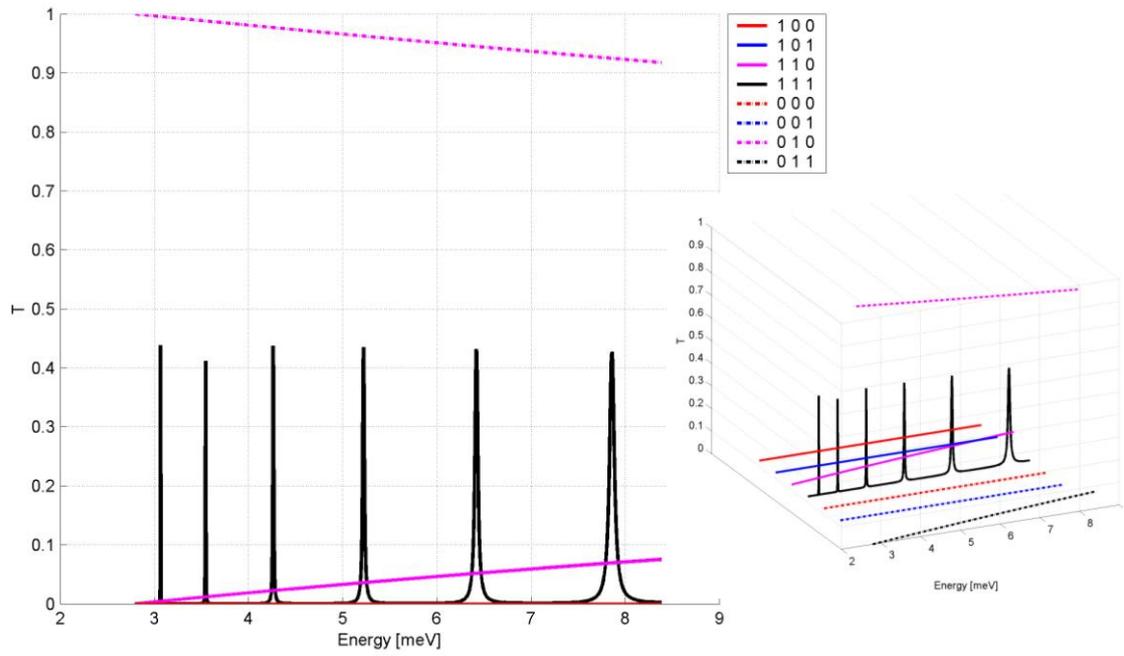

(c)

Fig. 2

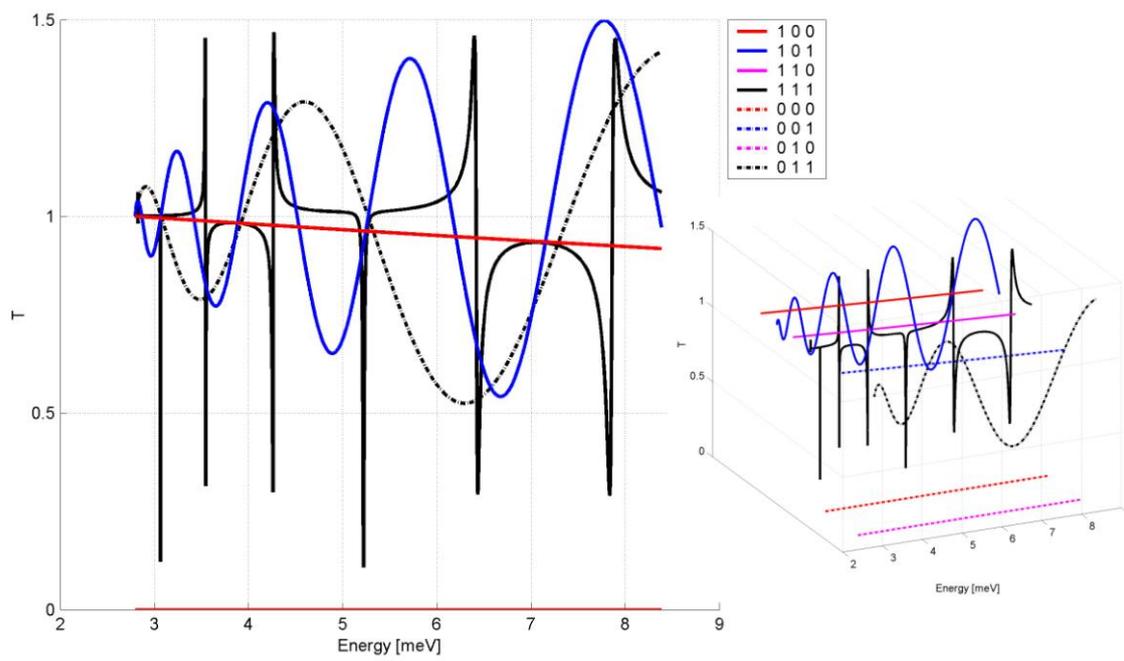

(a)

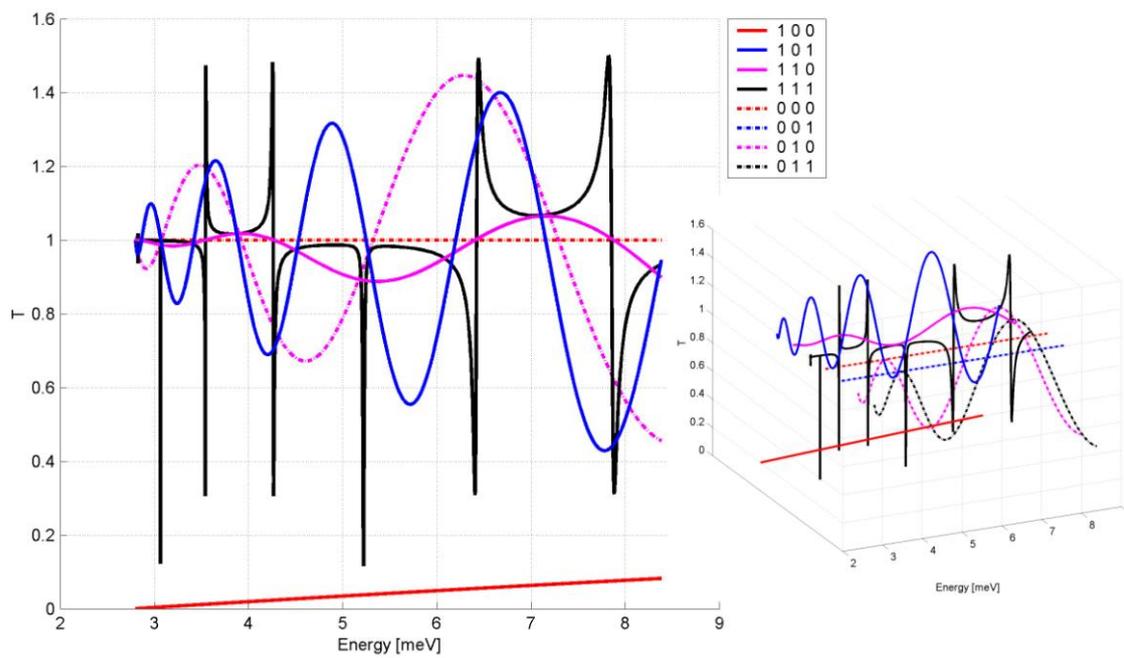

(b)

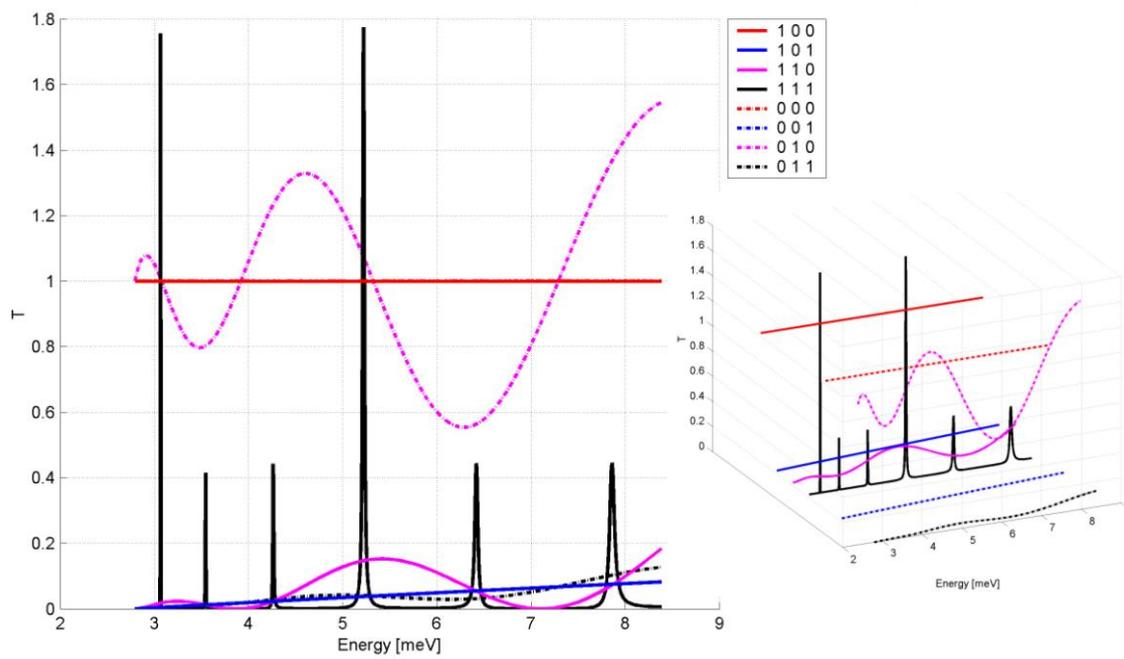

(c)

Fig. 3

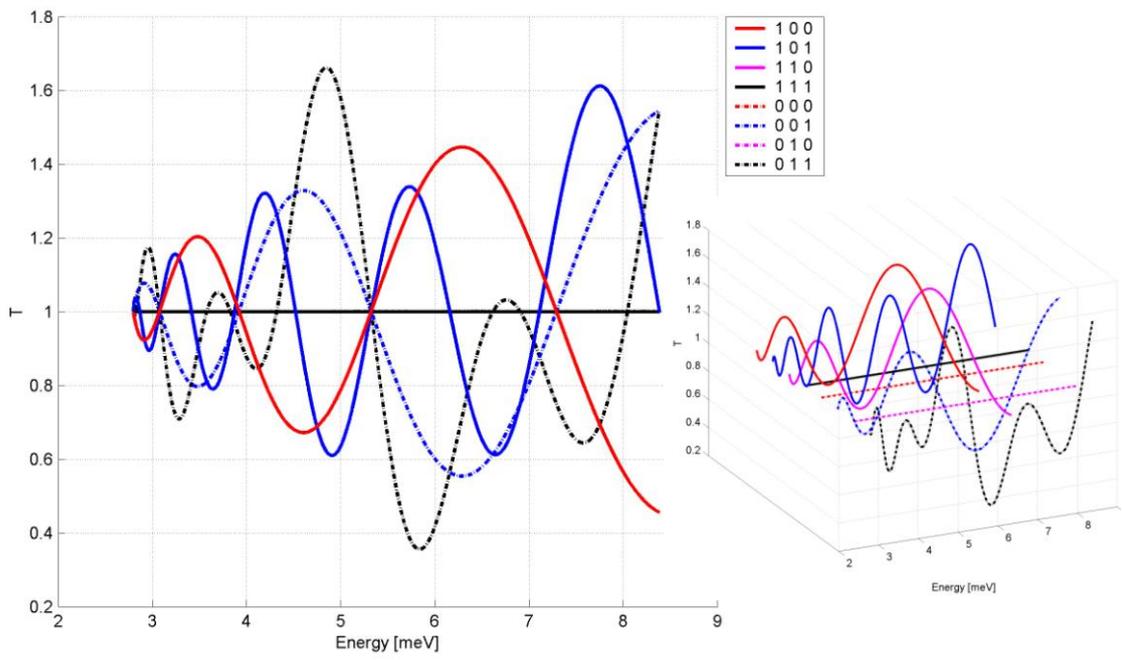

(a)

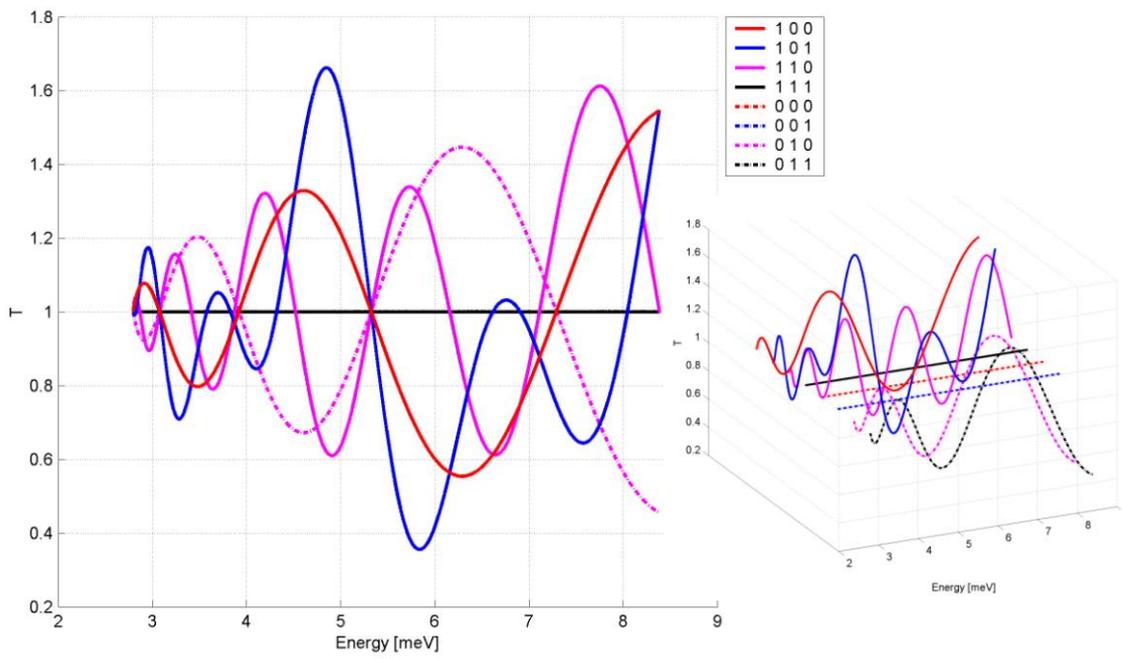

(b)

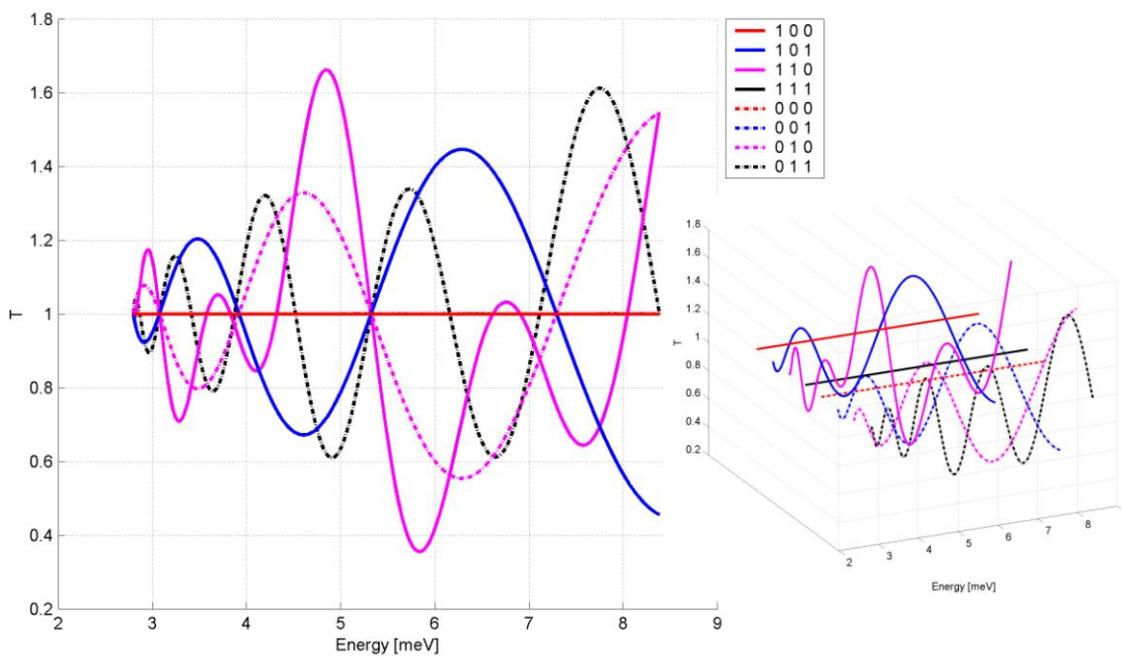

(c)

Fig. 4

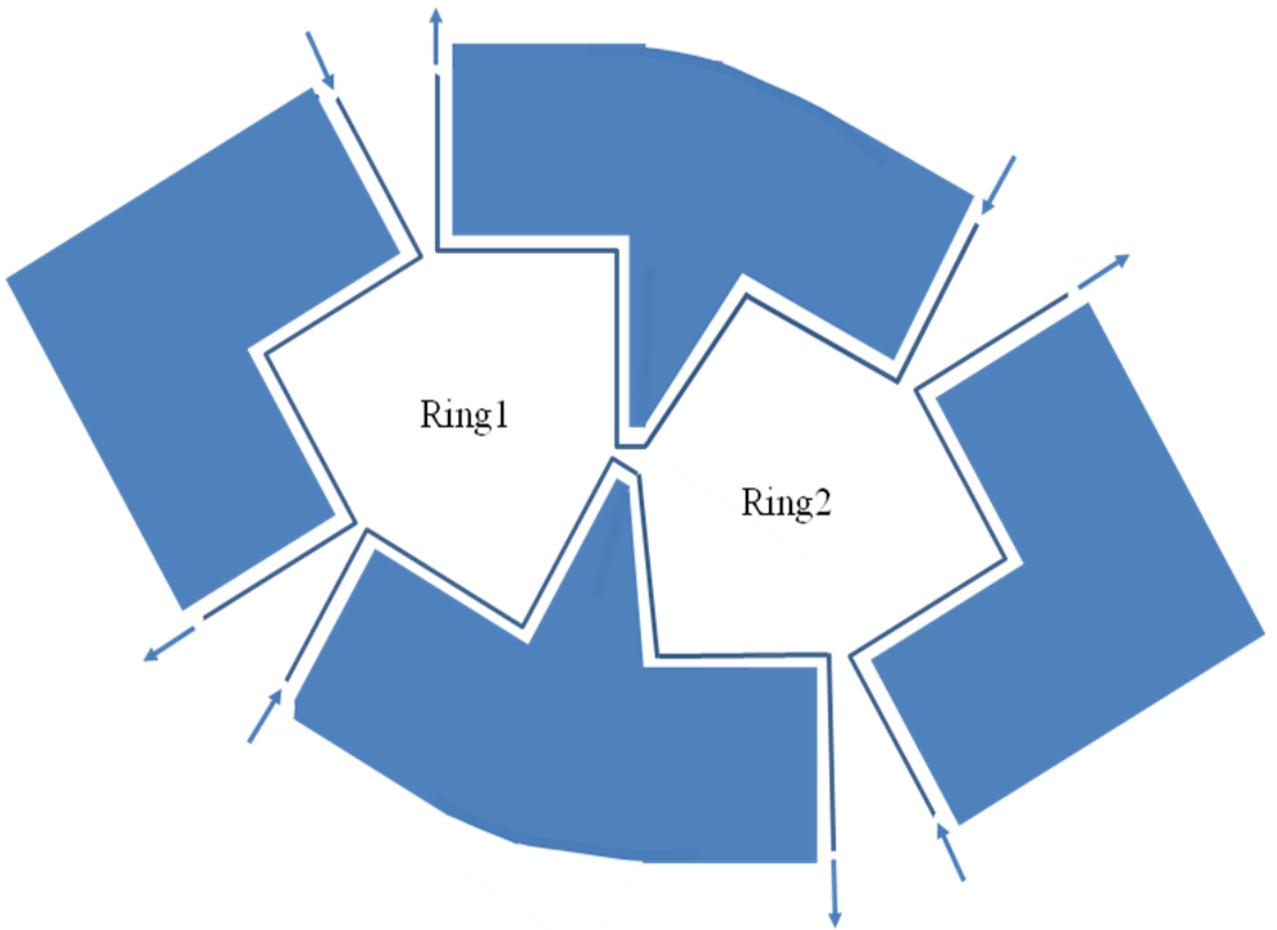

Fig. 5